\documentclass[10pt, twocolumn, conference]{IEEEtran}
\usepackage{xcolor}
\usepackage{scalerel} 
\usepackage{graphics}
\usepackage[utf8]{inputenc}
\usepackage{amsfonts}
\usepackage[fleqn]{amsmath}
\usepackage[hidelinks]{hyperref}
\usepackage{algorithm}
\usepackage{multirow}
\usepackage{booktabs}
\usepackage{amssymb}
\usepackage[flushleft]{threeparttable}
\usepackage{color}
\usepackage{array}
\usepackage{hhline}
\usepackage{xcolor,colortbl}
\usepackage{tablefootnote}
\usepackage{pifont}
\newcommand{\qed}{\nobreak \ifvmode \relax \else
      \ifdim\lastskip<1.5em \hskip-\lastskip
      \hskip1.5em plus0em minus0.5em \fi \nobreak
      \vrule height0.75em width0.5em depth0.25em\fi}

\ifCLASSINFOpdf
\else
\fi

\begin{document}

%\begin{singlespace}
\title{Preventing Distillation-based Attacks on Neural Network IP}

\author{\IEEEauthorblockN{Mahdieh Grailoo, Zain Ul Abideen, Mairo Leier and Samuel Pagliarini}
\IEEEauthorblockA{Centre for Hardware Security,  Dpt. of Computer Systems, Tallinn University of Technology (TalTech), Estonia \\
Email: \{\ mahdieh.grailoo, zain.abideen, mairo.leier, samuel.pagliarini \}\ @taltech.ee}\\\vspace*{-1.2cm}} 

\maketitle

\begin{abstract} 
Neural networks (NNs) are already deployed in hardware today, becoming valuable intellectual property (IP) as many hours are invested in their training and optimization. Therefore, attackers may be interested in copying, reverse engineering, or even modifying this IP. The current practices in hardware obfuscation, including the widely studied logic locking technique, are insufficient to protect the actual IP of a well-trained NN: its weights. Simply hiding the weights behind a key-based scheme is inefficient (resource-hungry) and inadequate (attackers can exploit knowledge distillation). This paper proposes an intuitive method to poison the predictions that prevent distillation-based attacks; this is the first work to consider such a poisoning approach in hardware-implemented NNs. The proposed technique obfuscates a NN so an attacker cannot train the NN entirely or accurately. We elaborate a threat model which highlights the difference between random logic obfuscation and the obfuscation of NN IP. Based on this threat model, our security analysis shows that the poisoning successfully and significantly reduces the accuracy of the stolen NN model on various representative datasets. Moreover, the accuracy and prediction distributions are maintained, no functionality is disturbed, nor are high overheads incurred. Finally, we highlight that our proposed approach is flexible and does not require manipulation of the NN toolchain.\end{abstract}

\begin{IEEEkeywords} Neural Network, Poisoning, Design Obfuscation, IP theft, Distillation.
\end{IEEEkeywords}

%--------------------------------------------------------------------
% -- INTRODUCTION (Section-1)
%--------------------------------------------------------------------
\section{Introduction} \label{sec:intro}
Systems based on Neural Networks (NNs) have experienced very rapid adoption in many application domains, including autonomous cars, facial recognition, surveillance, drones, and robotics \cite{rfHwAssDate,NN_app_2}. %NN_app_1, %\cite{rfdeepobfuscation}. 
In such systems, well-trained NN models (i.e., models that require significant time, resources, and effort to develop) are considered the owner's intellectual property (IP) \cite{rfHwAssDate}. Furthermore, since NNs are often considered for hardware (HW) acceleration, the risk of IP theft cannot be neglected. This is true for NNs implemented in both ASIC and FPGA forms. This risk is generally overlooked as most security concerns in NNs tend to relate to privacy \cite{zhang2015privacy} instead of IP ownership.

%In fact, since the fabless model of the integrated circuit (IC) ecosystem dictates that the entire circuit design must be shared for manufacturing, IP theft is a real threat. . In this case, in addition to IP theft for circuit reverse engineering or overproduction, the attacker's goal could also be to corrupt an IP by inserting malicious logic such as HW Trojans and malicious backdoors \cite{rfHWObfsIP}.

In order to protect hardware-based IPs, various obfuscation techniques have been proposed, including logic locking \cite{obs1}, IC camouflaging, split manufacturing, and their many variants. These techniques typically promote ``modifications'' at layout- or circuit-level to confuse and inhibit an adversary from reverse engineering the design. However, while the aforementioned techniques could be used to obfuscate virtually any type of IP, they offer an insufficient guarantee for the security of NN IPs. In particular, logic locking inserts extra logic into a circuit that locks its functionality behind a secret key that is stored in a tamper-proof memory. In the case of NN IPs, logic locking structurally obfuscates the NN circuit. However, logic locking is not necessarily an effective solution for the specific scenario of NN IPs. This is because frequently raised concerns about NN model piracy tend to refer to functionality theft \cite{rfdeepobfuscation}, i.e., theft of the weight parameters, and not of the circuit topology. In fact, the NN circuit topology is typically a published NN architecture with known high modeling capabilities \cite{rfHwAssDate}.

%Above all, logic locking is the most well-known for logic obfuscation which could also be exploited to hide a NN IP. 
 %\cite{obs1,logic_2}. %, logic_1, logic_2 %\cite{rfHWObfsIP,rfTchAlgObfs}. 
 
Alternatives to prevent the theft of well-trained NN models do exist and come in different forms. A recent approach towards obfuscation on NN IPs is proposed in \cite{rfHwAssDate}, where a key-dependent backpropagation algorithm is utilized to train the NN. However, this approach has limited practicality since its security hardness is tied to a tamper-proof memory and a specially crafted training algorithm, which implies a toolchain manipulation that is not trivial. Furthermore, contrary to logic locking schemes where an adversary is interested in breaking the key, an adversary might be interested in bypassing they key-based lock if the target circuit is a NN IP. This can be achieved  if an adversary has a functioning chip (``oracle'') to apply inputs to and get predictions out \cite{rfPrdPsn}. This type of oracle-guided attack is much more adequate for the problem at hand and, in fact, is a \textbf{distillation-based attack}. Such type of attack is the main concern of our paper. In this generous yet pragmatic white-box attack scenario, the NN architecture, the oracle, and the circuit netlist are all available to the adversary, making protecting an IP much harder.

\begin{figure*}[tb]
\centering
{\includegraphics[width=1.0\textwidth]{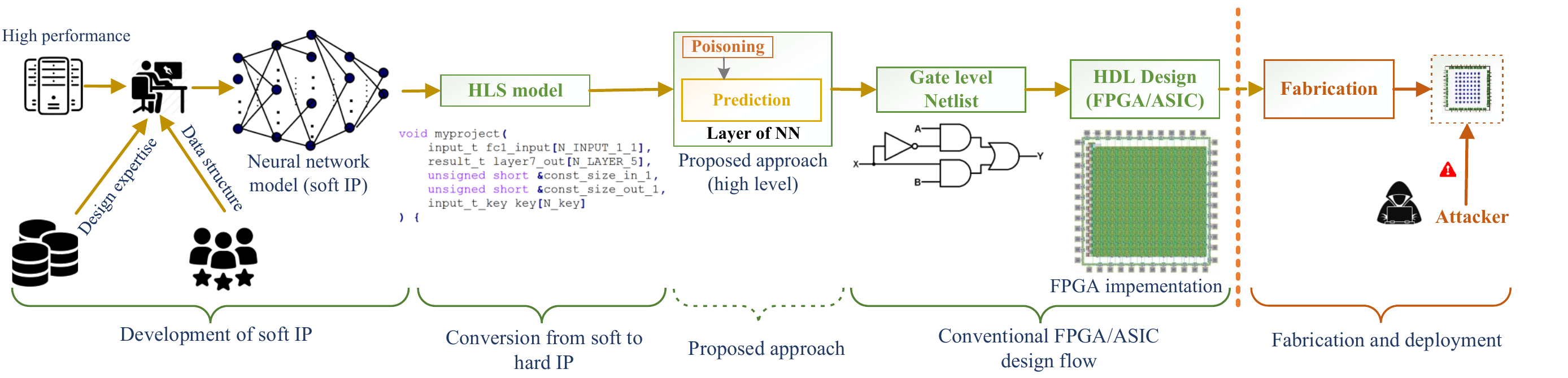}} \vspace{-10pt}
\caption{Design flow for obfuscation of NN IP, for both FPGA and ASIC designs.}
\label{figDF}
\end{figure*}

This paper then proposes techniques to poison the predictions of a NN which aim at reducing the threat of a distillation-based attack. Our method indirectly introduces noise in the predictions of a NN. Such noise is carefully crafted to maintain a high accuracy in the original model, whereas the stolen model will present a significant drop in accuracy. In summary, the contributions of this paper are as follows:

\begin{itemize}
    \item Use of an adequate threat model for attacks against NN IP that considers distillation. Ours is the first work to consider this scenario in NN IP obfuscation.
    \item Proposal of obfuscation methods that poison the predictions while incurring zero overheads. Our methods do not rely on tamper-proof memories or key-based locks.
    \item We perform no toolchain manipulation, such that the proposed obfuscation is compatible with an HLS-based design flow (see Fig. \ref{figDF}). This makes our methods agnostic to NN architectures and training processes, thus more practical and flexible than current approaches.
    \item Rich results for 5 different and representative neural network architectures.

\end{itemize}

%--------------------------------------------------------------------
% -- MOTIVATION AND THREAT MODEL (Section-2)
%-------------------------------------------------------------------

\section{Motivation and Threat Model} \label{sec:orctrb}
%In recent years, the NNs are being exploited for different applications, putting their security at stake. Numerous techniques have been proposed to obfuscate the circuit and they are still evolving with time. 

\begin{figure}[tb]
\centering
{\includegraphics[width=0.5\textwidth]{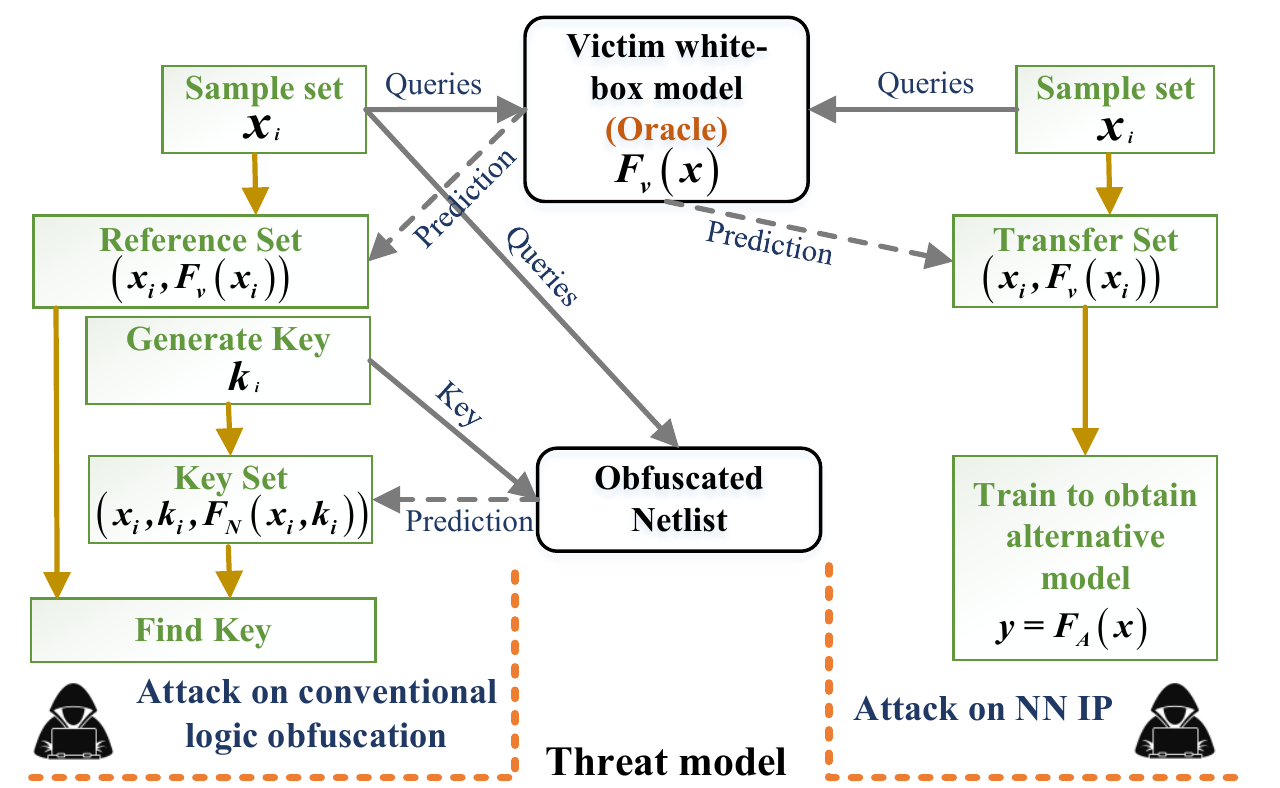}} \vspace{-10pt}
\caption{Contrast against different threat models.} \vspace{-10pt}
\label{figCMP}
\end{figure}

As we previously alluded to, security research for neural networks has often focused on privacy issues. On the other hand, hardware obfuscation is concerned with the theft of intellectual property. In order to clearly explain the threat we are concerned with, i.e., IP theft for NN hardware, a comparison of threat models is given in Fig. \ref{figCMP}.

%A generous but pragmatic attack scenario is considered where both the obfuscated netlist and an oracle are available, i.e., OG, that significantly complicates IP protection. Moreover, the attacker is aware of the NN architecture. the generic adversary's

It is essential to distinguish how obfuscating HW-implemented NNs is very different from random obfuscation of logic, regarding both adversary targets and assumptions. From the standpoint of IP's protection, the goal of an adversary is to obtain a ``victim" model $F_V(x)$. The victim model is, in fact, the IP to be protected. Notice how Fig. \ref{figCMP} describes two attacks: the attack on the left side is what the majority of works in the literature consider and is generally adequate for logic locking solutions. The attack described on the right side of the image is adequate for NN IP and one can appreciate how it is simpler, even visually.

The objective of the attacker on the left is to recover the key that was utilized to lock the design. As a customary, we suppose that the adversary has access to an oracle which can be queried. The adversary sends a query $x_i$ (input vector)\footnote{This is an input vector that is representative of the targeted dataset, e.g., a 28x28 image.} to the oracle and receives predictions resulting in a Reference Set that effectively is a mapping of inputs to outputs. The adversary also attempts to generate a key guess ($k_i$) and send the same $x_i$ to an obfuscated netlist model $F_N(x_i,k_i)$ to obtain a second set termed Key Set. With these two Sets (Reference and Key), the attacker executes well-known attacks as he/she tries to find the key for the obfuscated NN. In fact, the adversary utilizes clever observations of the prediction to exclude incorrect key guesses until a key is found for each every input-to-output mapping is satisfied. Prominent attacks \cite{rfsat, rfsmt, rfatpg} to recover the key from obfuscated circuits include Satisfiability (SAT), Satisfiability Modulo Theory (SMT), Sensitization/Automatic Test Pattern Generation (ATPG), and many others. Readers are directed to \cite{joeanalysis} for a detailed discussion on the modelling of such key-guessing attacks.

Next, we ought to explain the concept of distillation which will be exploited to formulate a distillation-based attack. The principal observation is that this process incurs (re)training of a neural network. If the attacker is aware of the network architecture, that gives him/her a formidable starting point for training. If the architecture is not known, the attacker may still try many different types of NNs and later the results from each training instance are combined into a final result. The process of transforming a complex model (i.e., the predictions from many heterogeneous NNs) into a simplified one is called distillation \cite{rfdistilling}. In simple words, distillation is a form of knowledge transfer between neural networks.

The right side of Fig. \ref{figCMP} illustrates the threat model we consider in this work. Concerning security, the adversary's target is to replicate the model $F_V$ by knowledge distillation. \emph{This is the first work to consider a distillation-based attack in the execution of HW obfuscation.} The attack consists of two steps: (i) query: the adversary uses the model $F_V$ as an oracle on a set of inputs to construct a ``transfer set" of input-prediction pairs $D^{Transfer} = {(x_i, F_V(x_i))}$; and (ii) training: the adversary trains an alternative NN using distillation  to minimize the empirical error on $D^{Transfer}$ \cite{rfPrdPsn}. The adversary is not required to break they key that protects the obfuscated netlist, nor is he/she required to generate the same number or value for the weights in $F_A$. It is also important to mention that the gradient-based optimization utilized in NN training takes less time than a brute force attack against $F_V$. 

%--------------------------------------------------------------------
% -- PROPOSED METHODOLOGY (Section-3)
%--------------------------------------------------------------------

\section{Secure Design Flow and Proposed Technique} \label{sec:VD}
%In practice, the semiconductor industry separates manufacturing (which occurs in foundries) from design (which occurs in fabless design houses). In this context, there is an increased risk of IP theft because the entire circuit design (layout) must be shared for fabrication with an untrusted party. In this scenario, the foundry itself may be a potential adversary interested in reverse engineering, overproduction, or IP manipulation. Moreover, even after ICs are fabricated and reach the market, they are still vulnerable to reverse engineering \cite{rfHWObfsIP,rfTchAlgObfs}. 

The complete design flow to secure the NN IP using our proposed methodology is shown in Fig. \ref{figDF}. This design flow is valid for both ASIC and FPGA implementations with the difference being that the ASIC fabrication is replaced by FPGA deployment. It consists of five different steps/stages where the obfuscation phase is highlighted with a dotted curly bracket. %Moreover, the stages are divided into trusted and untrusted, according to the fabless model of the IC ecosystem. 

A representative design flow for NNs consists of executing an ML toolchain, building the HLS model, HDL code, and drawing a layout or programming an FPGA. To this flow, we add the proposed approach for the obfuscation of NN IP using the poisoning of predictions. In the first stage, the designer develops a NN architecture based on the provided design criteria. He/she utilizes his/her designing expertise to make a NN IP well-structured and capable of high-performance. %\footnote{Usually, NN IP are deployed on sensitive and high-performance applications, therefore the designer keeps the basic principle of high-performance.}.  
In the second stage, the NN IP is typically converted from a high-level model to a C/C++ description \cite{rfhls4ml}. In the third stage, the NN IP is obfuscated with poisoning the predictions. In the fourth stage, a conventional HLS tool is used to map the C/C++ architecture into HDL. Once the HDL architecture is generated, then it can be easily implemented using conventional ASIC/FPGA flow. Once the ICs are fabricated/programmed, then they are deployed in the field. The attacker then attempts to steal the IP at this stage.

%We assume, without loss of generality, that the fabrication stage and the time when the IC reaches the field correspond to two threat models of oracle-less and oracle-guided. In an oracle-less attack, the attacker only has access to the obfuscated netlist. In an oracle-guided attack, the attacker has access to the netlist and a functional IC (i.e., the oracle) that is available on the open market \cite{rfthreatmodelOGOL}. Recalling again, these two threat models are generally well-understood for conventional logic-clocking, but they have not been sufficiently considered when NN models are obfuscated (see Section \ref{sec:orctrb}). 
%For example, HPNN \cite{rfHwAssDate} has neglected DbA in OG threat model.Then, the important weights from the most important layers are selected for obfuscation. The motivation for this is that an attacker cannot train the entire NN by knowledge distillation \cite{rfdistilling} and obtains the weights at the presence of oracle in the OG. The motivation here is that an attacker cannot obtain the obfuscated weights if he has the netlist from factory. Indeed, we have found that obfuscation of important weights is sufficient for IP protection of NNs. This is due to the fact that a further decrease in accuracy with more obfuscation encourages the attacker to bypass key discovery via a distillation-based attack \cite{rfdeepobfuscation}. 

\subsection{Obfuscation technique for NN: Poisoning the predictions} \label{sec:cad_flow}
The main goal of an attacker is to steal the model functionality based on the observed input-output pairs. The idea of defending against this attack is borrowed from software approaches that add perturbations while maintaining accuracy so that an attacker cannot train the NN entirely or partially. In \cite{rfknockoff}, the authors analyzed to what extent an adversary can steal the functionality of a NN software model based solely on black box interactions \cite{rfknockoff}. Other works, also in the software domain, have implemented distillation approaches to this end \cite{rfPrdPsn, rfrvrsgmd}. The existing software solutions against distillation attempt to constrain probability distributions or manipulate gradient direction during training. Another alternative in software is to use additional layers to actively balance the probability distributions to add ambiguity in the NN response. Unfortunately, translating these solutions directly to hardware would incur high overheads. However, we are interested in solutions that present either zero or low overheads. %Our proposed technique is software-inspired but carefully crafted to this end. 
%In this work, the authors analyzed to what extent, an adversary can steal the functionality of NN based solely on blackbox interactions \cite{rfknockoff}. This technique utilized the posterior distribution, which is set to zero to prevent the distillation-attack. A utility-constrained defense is proposed where the probabilities are poisoned such that the gradient direction actively maximizes the angular divergence between the original and poisoned gradient signals \cite{rfPrdPsn}. Another approach it to use an additional reverse-sigmoid layer which is added to the last layer to soften the posterior distribution and introduce ambiguity among the non argmax probabilities \cite{rfrvrsgmd}. 

 \label{sec:WS}
\begin{figure}[tb]
\centering
{\includegraphics[width=0.5\textwidth]{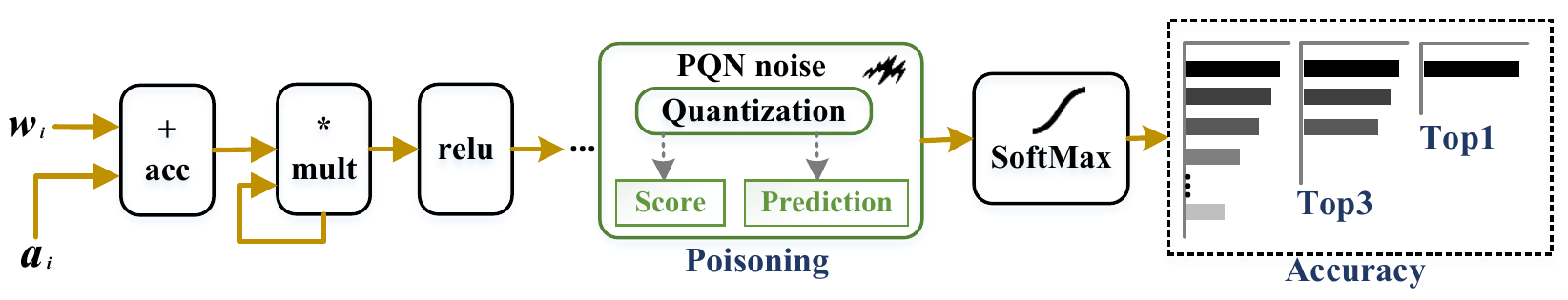}} \vspace{-10pt}
\caption{The proposed obfuscation of NN IP using poisoning the predictions.}\vspace{-10pt}
\label{figcir}
\end{figure}

Our obfuscation technique is mainly based on poisoning the predictions during the inference phase by inserting perturbations, as presented in Fig. \ref{figcir}. The presented perturbation does not destroy the distribution of predictions, thus hiding the presence of noise. Moreover, the initial class label of the model is maintained so as not to affect the accuracy. In other words, assuming the NN classifies input data into $n$ classes, we aim to maintain the prediction accuracy for the class with the highest score/probability while adding noise to the other classes. We aim to maintain the prediction order and probability distribution for the classes while adding noise to the probabilities. We use an idea based on the Pseudo Quantization Noise (PQN) model to generate perturbations with zero overhead. The PQN model is a practical way to handle quantization by modeling the process as additive stationary white noise that is uncorrelated over time \cite{rfqnbk}. The idea of quantization substitution with PQN is utilized oppositely to achieve a pseudo white noise perturbation via \textbf{truncation}. 

We perform truncation to both the end-result of the classification as well as to the intermediate score. We term these approaches \textbf{Prediction truncation (PT)} and \textbf{Score truncation (ST)}, respectively. Recalling again, this obfuscation is clearly different from the conventional logic obfuscation schemes like logic locking. Let's briefly discuss how we insert the truncation logic that performs quantization. Referring to Fig. \ref{figcir}, which is a generic representation of a typical NN architecture, notice that the quantization is performed around the SoftMax function. SoftMax is responsible for normalizing the output and converting it into probabilities. A modified SoftMax function that takes noise into account is given in Eq. \ref{sftmx}.

\begin{equation} \label{sftmx}
P(p=i|s,q)=\frac{e^{s_{i}+q_{i}}}{\Sigma_{j=1}^{C}e^{s_{j}+q_{j}}}
\end{equation}
where $s$ is a score vector whose length is equal to the number of classes $C$, and $q$ is the quantization noise. The $e$ in SoftMax is the exponential function, which significantly increases the probability of the largest score and decreases the probability of the smaller scores compared to the standard normalization. In ST, we manipulate the input of SoftMax directly (i.e., the score vector). In PT, we manipulate the output directly (i.e., the predictions). In either case, the relative probability distribution is maintained for accuracy reasons.

Still referring to Fig. \ref{figcir}, we also introduce an ``ArgMax'' truncation as a form of noise. In \textbf{Top1}, the highest probability value is kept while all others are set to zero. In \textbf{Top3}, the 3 highest values are kept. While this is effective in limiting the access an attacker has to the prediction values, it also reveals almost immediately that the ArgMax approach has been applied. Let us now discuss the advantages of the three truncation approaches with the aid of experimental results.

%--------------------------------------------------------------------
% -- EXPERIMENTAL RESULTS (Section-4)
%-------------------------------------------------------------------

\begin{figure*}[t]
\centering
{\includegraphics[width=0.9\textwidth]{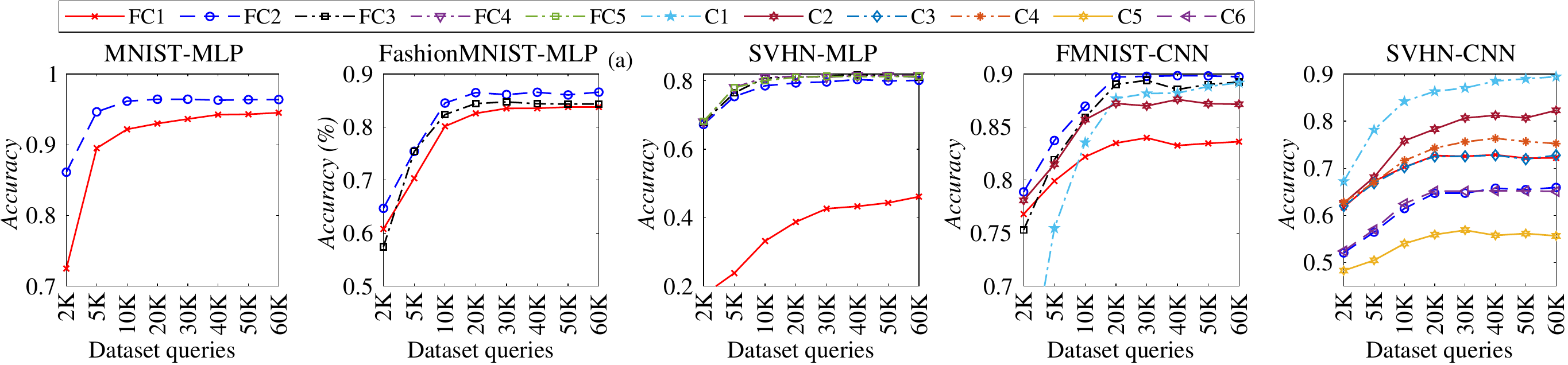}}\vspace{-10pt}
\caption{Accuracy per layer vs. dataset queries: A layer ranking for poisoning.}\vspace{-10pt}
\label{figspag}
\end{figure*}

\section{Experimental Results} \label{sec:results}
This paper evaluates the security analysis of five representative NN architectures. They are configured and trained on three popular image datasets for classification. These datasets include MNIST (MS), Fashion MNIST (FM) and SVHN (SV). We also utilize two classes of NNs, the Multi-layer perceptron (ML) and the convolutional neural network (CN). The HLS4ML tool is used to convert their Qkeras model into an HLS project \cite{rfhls4ml}. Aforementioned architectures with different datasets are trained using Adam (lr=0.0001) or SGD (lr=0.1) with Momentum (0.5) for 40 epochs. 

To start our analysis, we first evaluate the accuracy of different architectures over three datasets to establish a baseline. Table~\ref{tab:table_1} reports the accuracy of the non-obfuscated architectures, i.e., prior to applying our methods. The first column shows the name of the dataset and the type of architecture. The second column details the NN architectures and its sequence of layers. The third and fourth columns list the number of neurons per layer and the accuracy of the NN, respectively. One can observe that MLP offers the highest accuracy for MNIST and the lowest accuracy for the classification of SVHN. The CN networks consist of both fully connected and convolutions layers; thus, it is the second-highest candidate for the classification in terms of accuracy. 

\begin{table}[h]
\begin{threeparttable}
\caption{Accuracy of the NN architectures on different datasets  before executing the obfuscation\textsuperscript{\textbf{\textcolor{gray}{\ding{68}}}}.} \vspace{-10pt}
\label{tab:table_1}
\begin{tabular}{|p{1.cm}|p{2.65cm}|p{3.2cm}|p{0.45cm}|}
\hline
%\rowcolor{gray!10}
\textbf{Dataset}   & \textbf{Architecture} & \textbf{No. of neuron in layers} & \textbf{Acc.} \\ \hline
\textbf{MS-ML} & 2FC-1R-1S & 784-100-10 & 97.42 \\ \hline
\textbf{FM-ML} & 3FC-2R-1S & 784-(2)100-10 & 88.41 \\ \hline
\textbf{SV-ML} & 5FC-2R-1S & 3072-(4)200-10 & 81.38 \\ \hline
\textbf{FM-CN} & 2C-2A-3D-4R-3FC-1S & 784-32-64-256-128-10 & 90.90 \\ \hline
\textbf{SV-CN} & 6C-3M-3B-7R-2FC-1S  & 3072-(2)32-(2)64-(3)128-10 & 95.88 \\ \hline
\end{tabular}
\begin{tablenotes}
      \small
      \item [\textbf{\textcolor{gray}{\ding{68}}}] ML: Multi-layer perceptron, CN: convolutional neural network, MS: MNIST, FM: FashionMNIST, SV: SVHN, C: convolutional, M: max-pooling, A: average-pooling, FC: fully-connected layers, D: dropout, R: ReLu, S: SoftMax.
    \end{tablenotes}
  \end{threeparttable}
\end{table}

\subsection{Drawbacks of Logic Locking for NN Obfuscation} \label{drawbacks}

Let us assume an obfuscation approach for NNs in which the weights of a trained network are `hidden' behind a lock. With a mux-like structure, the lock selects between the original weight $w_i$ and its complement $-w_i$, as described in \cite{rfHwAssDate}. While this could be effective to hide the true weights, it implies the entire weight space is doubled, which incurs a heavy overhead. 

One clever observation to be made is that not all weights of a NN, even if very well trained, have the same importance. This is illustrated in Fig. \ref{figspag}. Each line corresponds to the accuracy of the network when all the weights of a single layer are obfuscated using a logic locking inspired scheme. As expected, the accuracy of an adversarial attack increases with the query budget\footnote{In order to quantify how many queries an attacker requires to steal the model, we introduce a notion of a query budget that is defined as the number of elements of the test set that the attacker has gained access to. From an attacker's point of view, the goal is to minimize this value. From a defence point of view, this number should be maximized.}, but not at the same rate for different layers. In other words, certain weights of certain layers are more important than others. From left to right, panels (a), (b) and (c) illustrate the results for three different datasets over MLP architecture. Similarly, panels (d) and (e) represent the trends line for the accuracy of CNN architecture.

It is noteworthy that the accuracy of a single layer in MLP's architecture shows an exponential behaviour. It starts increasing and it saturates for a large number of dataset queries. Concerning the FC1 layer in SVHN, it offers the least accuracy. Regarding panels (d) and (e), the accuracy trend is somewhat similar and tends to saturate with a large query budget, even if the relative accuracy of a single layer shows a more diverse pattern. For example, C1 offers the highest accuracy in panel (d). Moving further, it is possible to derive an ``importance ratio'' (R) from Fig. \ref{figspag}, which is calculated by dividing the accuracy of each layer by its number of parameters. This analysis is listed in Table. \ref{tab:table_2}.

%Considering the worst case, we select 60K dataset queries' accuracy and calculate layers' importance ratio against three datasets.  

\begin{table}[h]
\begin{threeparttable}
\caption{Importance ratio (R) for different NN architectures over three datasets under a 60K query budget\textsuperscript{\textbf{\textcolor{gray}{\ding{68}}}}.} \vspace{-10pt}
\label{tab:table_2}
\begin{tabular}{|p{0.8cm}|p{0.5cm}|p{0.61cm}|p{0.61cm}|p{0.61cm}|p{0.61cm}|p{0.61cm}|p{0.61cm}|}
\hline
\multirow{ 2}{*}{\textbf{Dataset}} & \multirow{ 2}{*}{\textbf{Type}} &\multicolumn{6}{c|}{\textbf{Importance Ratio (R)}}\\\cline{3-8}
&  & \textbf{\textit{1}}& \textbf{\textit{2}}&\textbf{\textit{3}}&\textbf{\textit{4}}&\textbf{\textit{5}}&\textbf{\textit{6}} \\\hline
\multirow{ 2}{*}{\textbf{MS-ML}} & F & 0.0012 & 0.0960 & N/A & N/A & N/A & N/A \\ \cline{2-8}
{} & {C} & N/A & N/A & N/A & N/A & N/A & N/A \\ \hline
\multirow{ 2}{*}{\textbf{FM-ML}} & F & 0.0010 & 0.0086 & 0.0843 & N/A & N/A & N/A \\ \cline{2-8}
{} & {C} & N/A & N/A & N/A & N/A & N/A & N/A \\ \hline
\multirow{ 2}{*}{\textbf{SV-ML}} & F & 0.0008 & 0.0020 & 0.0020 & 0.0020 & 0.0400 & N/A \\ \cline{2-8}
{} & {C} & N/A & N/A & N/A & N/A & N/A & N/A \\ \hline
\multirow{ 2}{*}{\textbf{FM-CN}} & F & 0.0003 & 0.0027 & 0.0690 & N/A & N/A & N/A \\ \cline{2-8}
{} & {C} & 0.1070 & 0.0017 & N/A & N/A & N/A & N/A \\ \hline
\multirow{ 2}{*}{\textbf{SV-CN}} & F & 0.0003 & 0.0500 & N/A & N/A & N/A & N/A \\ \cline{2-8}
{} & {C} & 0.1000 & 0.0087 & 0.0040 & 0.0020 & 0.0007 & 0.0004 \\ \hline
\end{tabular}
\begin{tablenotes}
      \small
      \item [\textbf{\textcolor{gray}{\ding{68}}}] F: importance ratio of fully-connected layer, C: importance ratio of convolutional layer.
    \end{tablenotes}
  \end{threeparttable}
\end{table}

Notably, the accuracy and R change from layer to layer. In the MLP networks, the importance of the layers increases from first to last. In the CNNs, the importance of the last fully connected layers and the first convolutional layer is the highest. It is clear from this analysis that, in general, the R values are rather small. The consequence is that a large amount of weights would have to be selected for any effective obfuscation to be achieved. Yet, these weights would have to be selectively chosen to keep overheads small. In our experiments, we have \textbf{not found} a compromise between these two requirements that would be reasonable and that would withstand a distillation attack. We recall again that logic locking techniques were not originally conceived for NNs. Hence, we turn our attention to the proposed method that promotes poisoning without manipulating the weights directly and without relying on a key or on a tamper-proof memory.

\subsection{Security Analysis of the Proposed Poisoning} \label{rslt}

In the security analysis that follows, the trained architectures are treated as an oracle that can be queried. The objective of an attacker remains to train $F_A$ to minimize the cross-entropy loss on a transfer set $D^{transfer}$. The attacker's accuracy is evaluated on the same test set, allowing a fair direct comparison with the accuracy of the original model. %Then, we proceed with the analysis of the obfuscation method against the distillation-based attack, and finally, we analyse the trade-offs between accuracy and dataset queries.

\begin{figure}[t]
\centering
{\includegraphics[width=1\linewidth]{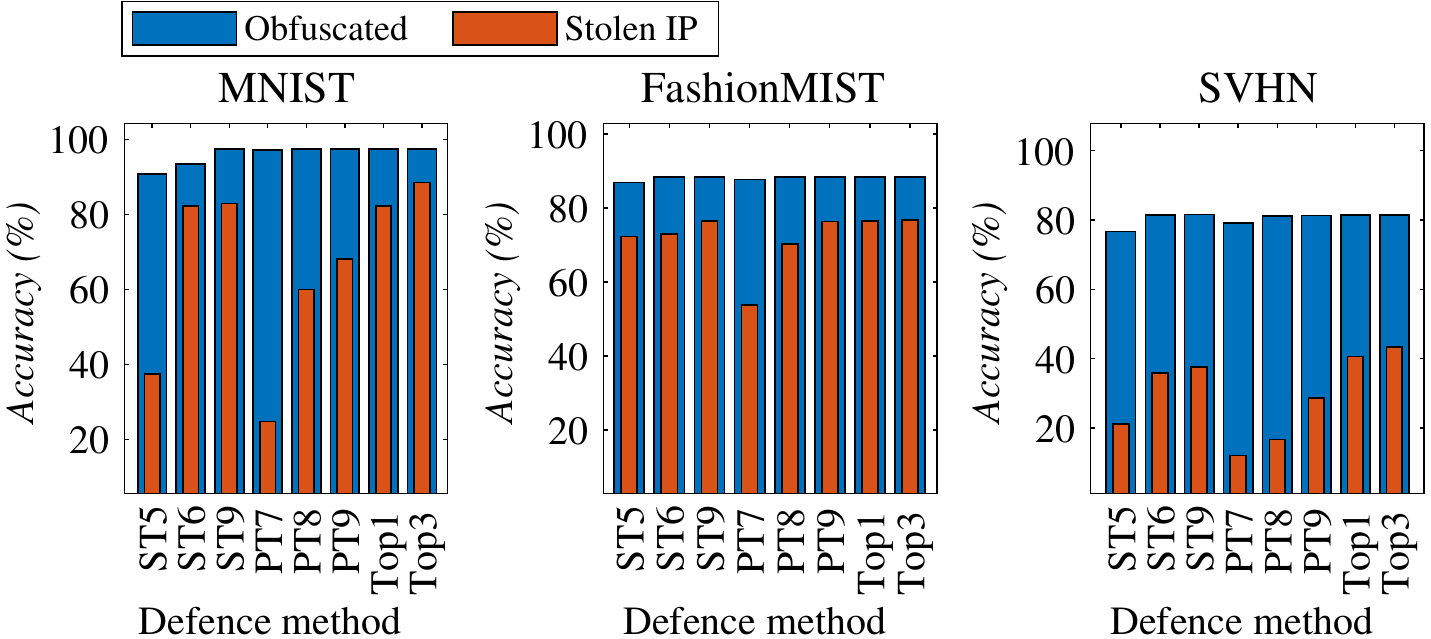}}\vspace{-10pt}
\caption{Summary of the effectiveness of the proposed defences.}\vspace{-10pt}
\label{oldplot}
\end{figure}

We have carried out a distillation-based attack on the aforementioned NN architectures. A summary of the effectiveness of the proposed defences is shown in Fig. \ref{oldplot}. The horizontal axis displays the different defence methods herein proposed. The vertical axis is the accuracy (\%) of the model, whereas dark grey bars correspond to the accuracy attained by the original owner of the obfuscated IP when using it. The light grey bars are the accuracy an adversary attains after a distillation-based attack is mounted. In this analysis, it is assumed that the attacker has a maximal query budget, i.e., he/she enjoys access to the entire dataset. The defense methods considered are ST (9-, 6-, and 5-bit), PT (9-, 8-, and 7-bit), Top1 (Top1), and Top3 (Top3) predicted labels. As seen, the stolen models in the Top1 and Top3 defences have the highest accuracy, which shows that the models can still be stolen. The ST defence successfully reduces the attacker's performance by at least 15\% across all datasets. This defence has a small impact on networks that consider simpler datasets (at least 15\% attacker's accuracy loss on MS-ML and FM-ML). In contrast, the impact is more significant on more complex datasets (at least 40\% accuracy drop for SV-ML).

\begin{figure}[tb]
\centering
{\includegraphics[width=1\linewidth]{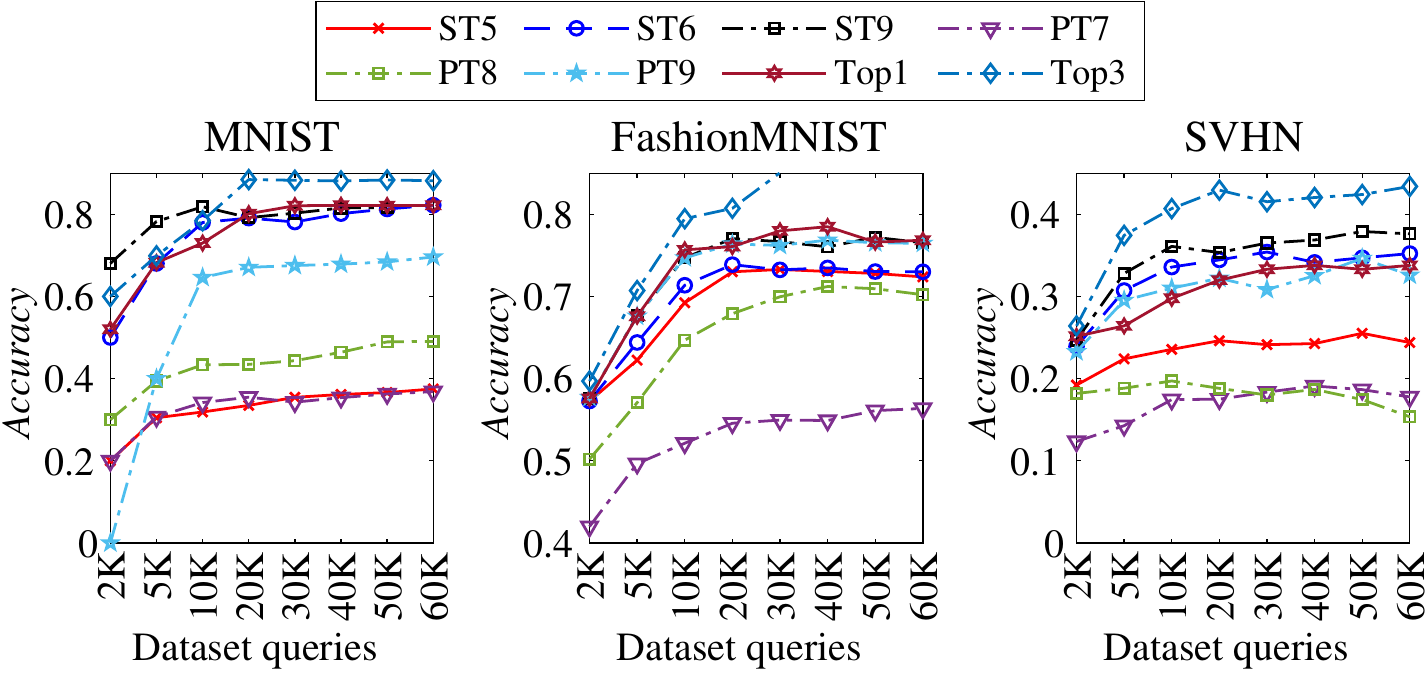}}\vspace{-10pt}
\caption{Evaluation of the accuracy of a model stolen via distillation-based attack versus attacker's query budget.}\vspace{-10pt}
\label{figAcc}
\end{figure}

The analysis in Fig. \ref{oldplot} is pessimistic since it assumes that the adversary has access to the entire dataset. This assumption is broken in the analysis portrayed in Fig~\ref{figAcc}, where the adversary query budget is varied. The accuracy of the stolen NN IP increases with the increase in the query budget size, and then it saturates at a certain point. This saturation narrows down if we move from a small dataset to a large dataset. The saturation for the MNIST dataset lies between 5K to 10K queries, between 2K to 5K for FashionMNIST, and less than 2K for SVHN. If we look into the right panel of Fig. \ref{figAcc}, the accuracy trend shows a slight variation. In a nutshell, all poisoning approaches appear to be effective for larger datasets. We selected only MLP to keep the same type of layers in architecture for this experiment. The predictions retain the probability distribution in ST, but most predictions become zero and have no distribution in PT, which visibly changes the NN functionality. Besides this, there are two other trends for Top1 and Top3. Using these methods, we observe that the saturation is sharp and offers a similar trend for FashionMNIST and SVHN datasets. This feature is of interest since an attacker may train multiple different types of NN on the same dataset during distillation (see Section \ref{sec:orctrb}).

In short, the classification accuracy of a stolen NN IP reduces for all obfuscation methods proposed. For both ST and PT, the attacker's accuracy is reduced by increasing the number of truncated bits. In particular, for the 5-bit ST, the attacker's accuracy decreases sharply but the original owner of the IP will observe a small (1-7\%) accuracy loss. More loss is observed for 7- and 8-bit PT due to a much higher probability bias. In fact, PT makes most predictions zero, entirely destroying the accuracy and functionality of the NNs. In Fig. \ref{figAcc}, the attacker's accuracy is plotted against the number of different attack queries to the defender. As already mentioned, increasing the query budget raises the attacker's accuracy, but the accuracy order for the defence methods remains the same as in Fig. \ref{oldplot}.   
As a result, although the attacker observes an accuracy loss in both truncations, ST provides non-replicability with significantly lesser perturbation. Therefore, \textbf{ST is recommended} as it \emph{gently} poisons the predictions and maintains the prediction distribution and accuracy. Notice that the ST light grey bars are nearly always above the PT light grey bars in Fig. \ref{oldplot}.

%PT provides non-replicability with significantly lesser perturbation. Therefore, \textbf{PT is recommended} as it \emph{gently} poisons the predictions and maintains the prediction distribution and accuracy. As mentioned earlier, the impact is more significant on larger datasets (at least 40\% accuracy drop for SVHN). Last but not least, the accuracy of Top1 and Top3 remains unchanged during the variation of truncated bits. But it offers a reasonable decrease in the accuracy of NN for larger datasets. This is closely dependent on the size of dataset queries. Therefore, it is not a practical solution, but it is combined with posing to aid in classifying larger datasets. %In fact, PT makes most predictions zero, entirely destroying the accuracy and functionality of the NNs.

%\begin{figure}[tb]
%\centering
%{\includegraphics[width=1\linewidth]{./Figures/Combined_three_scores_figure3.pdf}}\vspace{-10pt}
%\caption{Accuracy vs. truncation trend for both ST and PT along with accuracy of Top.}\vspace{-10pt}
%\label{figspag1}
%\end{figure}

\subsection{Performance-Power-Area (PPA) Overheads for the Proposed Poisoning} \label{subsubsec:hardware_cost} 
\begin{figure}[t]
\centering \footnotesize
\includegraphics[width=0.98\linewidth]{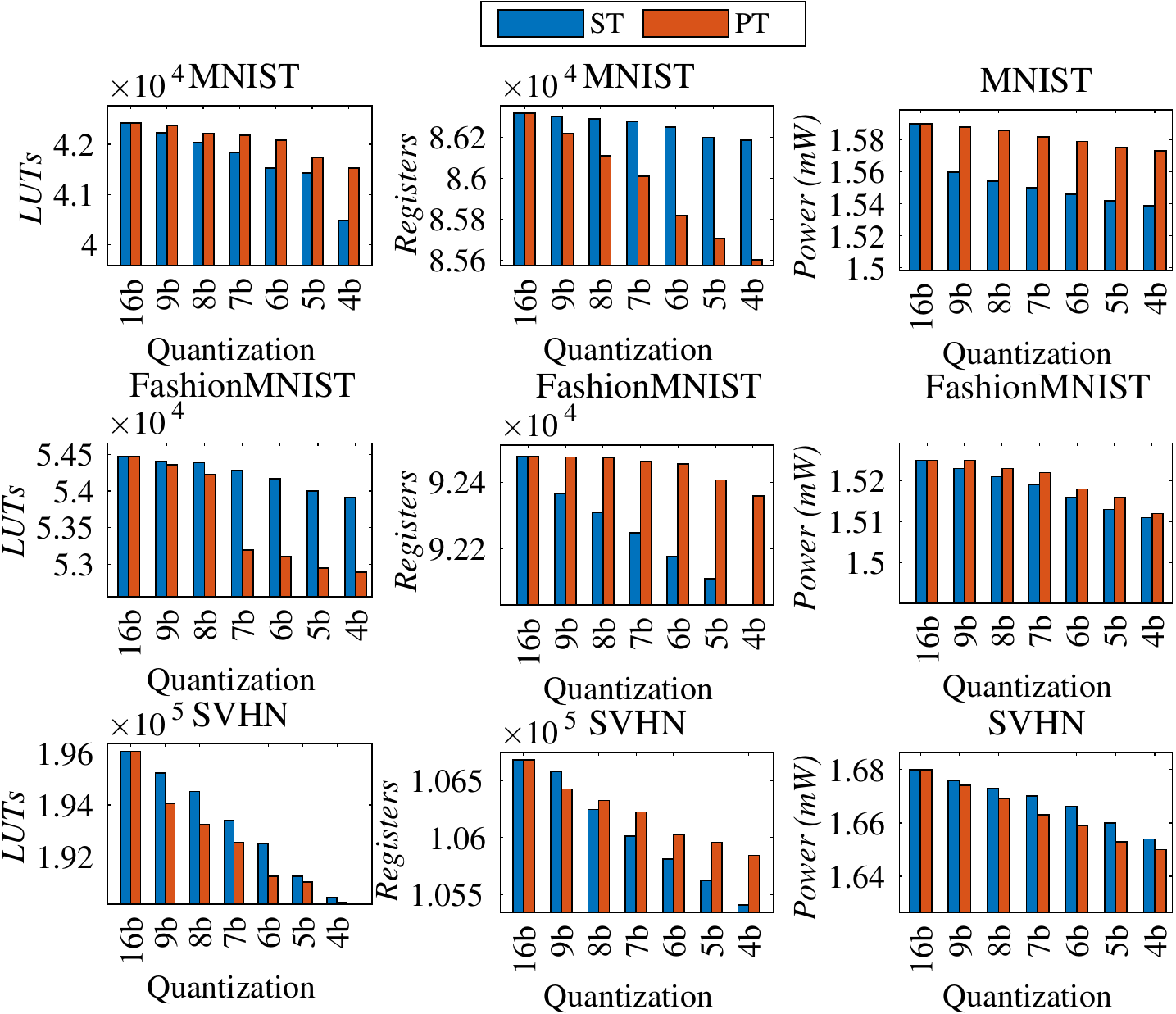}\vspace{-10pt}
\caption{PPA overheads for the proposed defenses ST and PTs for MLP architecture.}\vspace{-10pt}
\centering
\label{fig:a1}
\end{figure}

While our threat model and design flow are amenable to both FPGA and ASIC platforms, results are only given for an FPGA device. This is without loss of generality since we do not manipulate any BRAMs or DSPs, which are FPGA-specific elements that are not directly inferred in synthesized ASICs. The targeted device for FPGA synthesis is the Kintex-7 XC7K325T-2FFG900C for all the experiments. Fig. \ref{fig:a1} illustrates the results for defence methods (ST and PT) for various degrees of truncation. The x-axis shows the value of applied quantization and the y-axis represents the value of the corresponding parameter. To be clear, the non-obfuscated NN represents scores and predictions with 16 bits. In ST and PT, the overheads for LUTs, registers, and dynamic power illustrates a slight decrease for all datasets. These overheads are not significant due to the truncation that happens in the last layer followed by SoftMax. With the increase in truncation, there is a slight decrease in LUTs, registers and power for MLP architecture. The frequency, static power, number of BRAMs, and DSPs remain the same for all circuits in each dataset. Considering the Top1 and Top3 methods, they present a different behaviour. The variation in quantization does not affect resources as there is just a slight increase in LUTs and registers ($<$1\%). Both MNIST and FashionMIST offer 100MHz, but SVHN is slightly slower and offers 83 MHz. Last but not least, the frequency of the obfuscated NN remains unchanged for ST, PT, Top1, and Top3.
%We have designed Tp1 to select the maximum accuracy. %ArgMax is placed at the end of SoftMax to hide all the prediction values and just pass the maximum one. This limits the access of an attacker to the prediction values for the oracle-based attacks. 

%--------------------------------------------------------------------
% -- COMPARISON AND DISCUSSION (Section-5)
%-------------------------------------------------------------------

\section{Comparison and discussion} \label{sec:comparison_discussion}
This work has proposed a practical obfuscation method for NN IPs that contrasts sharply with current obfuscation approaches. The proposed protection accommodates the HLS-based design flow for hardware-implemented NNs without requiring any manipulation of the NN toolchain or access to the NN model. Our method also contrasts with other approaches requiring toolchain manipulation and a complete understanding of an NN architecture, limiting their practical application. Software solutions, if and when ported to hardware, would suffer from high overheads. On the other hand, our solutions do not infer any extra overhead for the obfuscation. Our method also incurs a zero clock cycle delay overhead. This is also true for \cite{rfHwAssDate}, however, the oracle-guided threat model is not considered, which may skew their attack resiliency. In comparison, our poisoning strategies have similar (or smaller) overheads and discourage oracle-guided attacks. Finally, software methods can be exploited along with our approach \cite{rfPrdPsn,rfrvrsgmd}. This possibility remains as future work. Furthermore, a direct comparison with logic locking schemes is not performed since we have extensively discussed how inappropriate they are when it comes to obfuscating NN IPs.

%--------------------------------------------------------------------
% -- CONCLUSION (Section-6)
%-------------------------------------------------------------------

\section{Conclusion}
In this paper, we have proposed a novel method for obfuscating the NN IP to prevent IP theft. We have proposed a practical obfuscation method that poisons the predictions and does not require a key to attain full functionality. Our approach is key-less and it is independent of the number of weights. It accommodates the obfuscated HLS model in the hardware design flow for NNs without requiring any manipulation of the NN toolchain or access to the NN model. The proposed model enjoys low overhead and, for the first time, considers the distillation-based attack in an oracle-guided threat model. Our results confirm a massive decrease in the classification accuracy over different datasets. 
%The proposed model reduces the attacker's performance between 15\% and 60\% over all datasets requiring tens of thousands of queries. 

\section{Acknowledgements}
\thanks{This work was partially supported by the EC through the European Social Fund in the context of the project ``ICT programme".}

\bibliographystyle{IEEEtran}
\bibliography{obfuscation}
\end{document}